\documentclass[12pt,aps,preprint,nofootinbib]{revtex4}

\usepackage{epsfig}
\usepackage{amssymb}
\usepackage{amsmath}
\usepackage{amsfonts}
\usepackage{graphics}
\usepackage{axodraw}

\newcommand{\mtrx}[2]{\left(\begin{array}{#1} #2 \end{array}\right)}

\newcommand{\eps}{\varepsilon}
\newcommand{\Refs}{Refs.}
\newcommand{\Ref}{Ref.}

\newcommand{\tab}{Tab.}
\newcommand{\eq}{Eq.}
\newcommand{\fig}{Fig.}

\newcommand{\bld}[1]{\boldsymbol{#1}}

\newcommand{\ie}{\emph{i.e.}}
\newcommand{\eg}{\emph{e.g.}}

\DeclareMathOperator{\real}{Re}

\allowdisplaybreaks

\preprint{MPP-2009-83}
\preprint{EURONU-WP6-09-05}

\begin{document}

\title{~ \\General bounds on non-standard neutrino interactions}

\author{Carla Biggio}
\email[]{biggio@mppmu.mpg.de}
\author{Mattias Blennow}
\email[]{blennow@mppmu.mpg.de}
\author{Enrique Fern\'andez-Mart\'\i nez}
\email[]{enfmarti@mppmu.mpg.de}
\affiliation{Max-Planck-Institut f\"ur Physik \\ (Werner-Heisenberg-Institut) \\ F\"ohringer Ring 6, 80805 M\"unchen, Germany\\}

\begin{abstract}
We derive model-independent bounds on production and detection
non-standard neutrino interactions (NSI). We find that the constraints
for NSI parameters are around $\mathcal O(10^{-2})$ to $\mathcal
O(10^{-1})$. Furthermore, we review and update the constraints on
matter NSI. We conclude that the bounds on production and detection
NSI are generally one order of magnitude stronger than their matter
counterparts.
\end{abstract}

%\keywords{neutrino oscillations, non-standard interactions}

\maketitle

\section{Introduction}

The formalism of non-standard neutrino interactions (NSI) is a very
widespread and convenient way of parametrising the effects of new
physics in neutrino
oscillations~\cite{Wolfenstein:1977ue,Mikheev:1986gs,Roulet:1991sm,Guzzo:1991hi,Brooijmans:1998py,GonzalezGarcia:1998hj,Bergmann:2000gp,Guzzo:2000kx,Guzzo:2001mi}.
Even though present data constrain NSI to be a subleading effect in
neutrino oscillation experiments, the possibility of their eventual
detection or interference with neutrino oscillations at
present~\cite{Grossman:1995wx,Ota:2002na,Friedland:2005vy,Kitazawa:2006iq,Friedland:2006pi,Blennow:2007pu,EstebanPretel:2008qi,Blennow:2008ym}
and
future~\cite{GonzalezGarcia:2001mp,Gago:2001xg,Huber:2001zw,Ota:2001pw,Campanelli:2002cc,Blennow:2005qj,Kopp:2007mi,Kopp:2007ne,Ribeiro:2007ud,Bandyopadhyay:2007kx,Ribeiro:2007jq,Kopp:2008ds,Malinsky:2008qn,Gago:2009ij}
experiments has triggered a considerable interest in the community.
In particular, it is a common practice to study NSI in matter, which
correspond to neutral-current-like operators, assuming that the
constraints on the NSI affecting production and detection processes
are much stronger.  However, up to now, only model-dependent bounds on
such interactions are present in the literature
\cite{Grossman:1995wx,GonzalezGarcia:2001mp}. The main aim of this
paper is filling this gap and providing model-independent bounds on
NSI affecting neutrino production and detection processes. Also, the
present constraints on matter NSI will be reviewed and updated.

Before entering into details and deriving the current bounds, a
discussion on the naturalness of large NSI is in order. In particular,
this argument has been faced for matter
NSI~\cite{Antusch:2008tz,Gavela:2008ra,Biggio:2009kv}, but the main
message can be applied to production and detection NSI as well.
Matter NSI are defined through the following addition to the
Lagrangian density:
\begin{equation}
 \mathcal L^{M}_{\rm NSI} = -2\sqrt{2} G_F \eps^{fP}_{\alpha\beta} 
[\bar f \gamma^\mu P f][\bar\nu_\alpha \gamma_\mu P_L \nu_\beta],
\label{matterNSI}
\end{equation}
where $f = e, u, d$ and $\eps^{fP}_{\alpha\beta}$ encodes the
deviation from standard interactions. For example, an operator of this kind is induced in fermionic seesaw models once the heavy fermions (singlets or triplets) are integrated out leading to a $d=6$ operator that modifies the neutrino kinetic energy \cite{DeGouvea:2001mz,Broncano:2002rw,Antusch:2006vwa,Abada:2007ux}. After a transformation to obtain canonical kinetic terms, modified couplings of the leptons to the gauge bosons, characterized by deviations from unitarity of the leptonic mixing matrix, are induced. Upon integrating out the gauge bosons with their modified couplings, NSI operators are therefore obtained. Because of
the strong bounds on the unitarity of this matrix, these NSI are
constrained to be $\lesssim
\mathcal{O}(10^{-3})$~\cite{Tommasini:1995ii,Antusch:2006vwa,Antusch:2008tz}. This
means that their eventual detection is challenging, although not
impossible, at future
facilities~\cite{FernandezMartinez:2007ms,Holeczek:2007kk,Goswami:2008mi,Altarelli:2008yr,Antusch:2009pm}.

On the other hand, large NSI could be generated by some other new
physics, not necessarily related to neutrino masses, at an energy
above the electroweak scale. As a consequence, an $SU(2)$ gauge
invariant formulation of NSI is mandatory. The simplest gauge
invariant realization of the operator in \eq~(\ref{matterNSI}) implies
to promote the neutrino fields to full lepton doublets. However, in
that case, strong bounds stemming from four-charged-fermion processes
would apply~\cite{Berezhiani:2001rs,Ibarra:2004pe,Raidal:2008jk}. In
order to avoid these constraints, cancellations among different
higher-dimensional operators are
required~\cite{Berezhiani:2001rs,Gavela:2008ra}. In the case of $d=6$
operators there is only one combination which satisfies these
conditions and the corresponding NSI are also severely
constrained~\cite{Cuypers:1996ia,Antusch:2008tz}. In the case of $d=8$
operators it has been shown that, avoiding cancellations between
diagrams involving different messenger fields or the introduction of
new leptonic doublets that could dangerously affect the electroweak
precision tests, the only possibilities of evading the constraints
imposed by gauge invariance reduce to the cases already mentioned,
with the consequent stringent bounds~\cite{Antusch:2008tz}. Therefore,
in order to realise the cancellations that would allow large NSI, some
fine-tuning is needed. An example of the naturalness prize required is
presented in the toy model proposed in
\Ref~\cite{Gavela:2008ra}. However, even if large NSI are generated in
this way at tree-level, dangerous quadratic divergences contributing
to four-charged-fermion operators appear at
one-loop~\cite{Biggio:2009kv}. In order to have large NSI, another
fine-tuning would then be required at one-loop unless the scale of new
physics is smaller than $4 \pi v$, where $v$ is the Higgs
vev. Alternatively, a symmetry could guarantee the cancellation both
at tree- and loop-level, but so far no model has been found with these
characteristics, \ie, leading to large NSI.

From the previous discussion it is clear that it is not easy to induce
large neutrino NSI in a specific theoretical framework. However, since
it is impossible to exclude them completely in a model-independent way
and since their effects may be visible at future experiments,
%giving precious hints on the nature of the underlying new physics, 
we think it is worthwhile to derive their present bounds.

The rest of the paper is organized as follows: First, in
Section~\ref{sec:cc}, we define charged-current-like NSI and derive
bounds on various combinations of $\eps$, specifying which bounds can
be set if only one non-zero $\eps$ is considered at a time.  We then
proceed by discussing loop bounds on charged-current-like NSI in
Section~\ref{sec:loops}. Finally, we review and update the bounds on
matter NSI in Section~\ref{sec:nc} and make a summary of the results
and conclude in Section~\ref{sec:results}.

\section{Charged-current-like non-standard interactions}
\label{sec:cc}

Let us start by considering NSI for source and detector
processes. Since these are always based on charged-current processes
so as to tag the neutrino flavour through the flavour of the
associated charged lepton, we will refer to these as
charged-current-like NSI. The general leptonic NSI are given by the
effective Lagrange density
\begin{equation}
 \label{eq:leptonCCNSI}
 \mathcal L_{\rm NSI}^\ell =
 -2\sqrt{2} G_F \eps^{\alpha\beta P}_{\gamma\delta}
 [\bar\ell_\alpha \gamma^\mu P \ell_\beta] 
 [\bar\nu_\gamma \gamma_\mu P_L \nu_\delta],
\end{equation}
where $P$ is either $P_L$ or $P_R$ and, due to Hermiticity,
$\eps^{\alpha\beta P}_{\gamma\delta} = \eps^{\beta\alpha
  P*}_{\delta\gamma}$. For charged-current-like NSI $\alpha \neq \beta$; in particular,
$\alpha =\mu$ and $\beta = e$ are the only parameters of importance for
neutrino oscillation experiments due to their effect in neutrino production via muon
decay. Notice that $\alpha = \beta = e$ would instead correspond to matter NSI. 

In a similar fashion, the charged-current-like NSI with quarks are
given by the effective Lagrange density
\begin{equation}
 \mathcal L^q_{\rm NSI} = -2\sqrt{2} G_F \eps^{qq' P}_{\alpha\beta} V_{qq'} 
 [\bar q \gamma^\mu P q'][\bar\ell_\alpha \gamma_\mu P_L \nu_\beta] + {\rm h.c.},
\label{eq:quarkCCNSI}
\end{equation}
where $q$ is an up-type and $q'$ is a down-type quark. Naturally, only
$q = u$ and $q' = d$ are of practical interest for neutrino
oscillations, due to their contributions to charged-current
interactions with pions and nuclei. Because of this, we will
concentrate on constraining $\eps^{\mu e}_{\alpha\beta}$ as well as
$\eps^{ud}_{\alpha\beta}$. Since the relevant combinations of NSI that
contribute to some processes will be of an axial or vector structure
we define
\begin{eqnarray}
\eps^{\gamma \delta V}_{\alpha \beta} &=& \eps^{\gamma \delta R}_{\alpha \beta} + \eps^{\gamma \delta L}_{\alpha \beta}, \\
\eps^{\gamma \delta A}_{\alpha \beta} &=& \eps^{\gamma \delta R}_{\alpha \beta} - \eps^{\gamma \delta L}_{\alpha \beta},
\end{eqnarray} 
in order to simplify the notation. Notice that more general Dirac
structures such as scalar or tensor couplings can in principle be
considered to generalise Eqs.~(\ref{eq:leptonCCNSI}) and
(\ref{eq:quarkCCNSI}). However, these NSI will have the wrong
chirality to contribute coherently with the SM production and
detection processes --something that is usually assumed for NSI-- and
therefore linear interference of these NSI will require an extra
chirality suppression~\cite{Kopp:2007ne,Delepine:2009am}. For this reason we will
neglect them here.

\subsection{Bounds from kinematic Fermi constant}

At present, the most precise determination of the Fermi constant $G_F$
is through the muon decay rate. However, if NSI of the form $\eps^{\mu
  e}_{\alpha\beta}$ are present, this will make the measured Fermi
constant from muon decays $G_\mu$ differ from the true Fermi constant
according to the relation $G_\mu = G_F f(\eps^{\mu e L}_{e
  \mu},\sum_{\alpha\beta P} |\eps^{\mu e P}_{\alpha\beta}|^2)$. Here
we have introduced the function
\begin{equation}
f(x,y) = 1 + 2 \real (x) + y,
\end{equation}
where $x$ represents the interference between the SM and the
particular NSI that contributes coherently with the SM to the process
and $y$ is the incoherent sum of the NSI contributions. Therefore, in
all the processes considered, stronger bounds will be implied for the
real part of $x$. Given the relation between $G_\mu$ and $G_F$, an
independent measurement of the Fermi constant will constrain
$\eps^{\mu e P}_{\alpha\beta}$. We will consider two different ways of
deriving the value of $G_F$, one involving only the kinematic
measurements of the gauge boson masses and one involving comparison to
the quark sector.

For determining $G_F$ from kinematic considerations, we need to review
the predictions of the Standard Model. From \Ref~\cite{Amsler:2008zz},
we have
\begin{equation}
 M_W = \frac{A_0}{s_W \sqrt{1-\Delta r}},
\end{equation}
where $A_0 = \sqrt{\pi \alpha/(\sqrt{2}G_F)}$, $s_W^2 = 1-M_W^2/M_Z^2$,
$\alpha$ is the fine-structure constant, and $\Delta r = 0.03690 \pm
0.0007$ is the radiative correction
to the tree-level relation. Thus, we obtain the relation
\begin{equation}
 G_F = \frac{\pi\alpha M_Z^2}{\sqrt 2 M_W^2(M_Z^2-M_W^2)(1-\Delta r)}.
\end{equation}
For the masses of the vector bosons, we use the combined fit for the
$W$ mass from LEP and Tevatron, $M_W = 80.398\pm 0.025$~GeV, as well
as $M_Z = 91.1876 \pm 0.0021$~GeV from LEP~\cite{Amsler:2008zz}. The
resulting Fermi coupling constant is
\begin{equation}
 G_F = (1.1696 \pm 0.0020)\cdot10^{-5} \ {\rm GeV^{-2}} \quad (1\sigma).
\end{equation}
Comparing with $G_\mu$, we obtain
\begin{equation}
\frac{G_\mu}{G_F} = 
f(\eps^{\mu e L}_{e \mu},\sum_{\alpha\beta P} |\eps^{\mu e P}_{\alpha\beta}|^2) = 
\frac{1.16637\pm0.00001}{1.1696\pm0.0020} = 
0.9973 \pm 0.0017,
\end{equation}
which represents a 90~\% confidence level agreement with the Standard
Model expectation.  The only truly model-independent bound that we can
extract from this is on the combination $f(\eps^{\mu e L}_{e
  \mu},\sum_{\alpha\beta P} |\eps^{\mu e P}_{\alpha\beta}|^2)$.  On
the other hand, it is common practice to assume the presence of only
one non-zero $\eps$ at a time in order to avoid cancellations inside
$f(\eps^{\mu e L}_{e \mu},\sum_{\alpha\beta P} |\eps^{\mu e
  P}_{\alpha\beta}|^2)$. In this way, the following bounds can be
obtained:\footnote{Throughout the paper we will follow the statistical
  approach proposed in \Ref~\cite{Feldman:1997qc} by Feldman and
  Cousins.}
\begin{eqnarray}
\real(\eps^{\mu e L}_{e\mu}) &=& (-1.4\pm1.4)\cdot10^{-3},\\
|\eps^{\mu eP}_{\alpha\beta}| &<& 0.030,
\end{eqnarray}
at 90~\% confidence level.

\subsection{Bounds from CKM unitarity}

One way of constraining the completely leptonic NSI, as well as some
of the charged-current NSI with quarks, is to make the assumption that
the Cabibbo--Kobayashi--Maskawa (CKM) matrix is unitary, as predicted
by the Standard Model. The experimental test of the CKM unitarity is
essentially based upon the determination of $V_{ud}$ and $V_{us}$ from
beta- and Kaon-decays,\footnote{In principle $V_{ub}$ should also be
  considered. However, its value is smaller than the uncertainty in
  the other two matrix elements and we therefore leave it out of our
  discussion.} where the Fermi constant extracted from muon decay
$G_\mu$ is used to predict the decay rates. These are proportional to
\begin{equation}
  \Gamma \propto G_F^2 |V_{ux}|^2,
\end{equation}
which means that, by inserting $G_\mu$ in place of $G_F$, we are
actually determining $|V_{ux}^M|^2 \equiv |V_{ux}|^2/f^2(\eps^{\mu e
  L}_{e \mu},\sum_{\alpha\beta P} |\eps^{\mu e
  P}_{\alpha\beta}|^2)$. Adding the information from beta- and
Kaon-decay experiments and assuming that leptonic NSI dominate over
quark NSI, we have~\cite{Amsler:2008zz}
\begin{equation}
 |V_{ud}^M|^2 + |V_{us}^M|^2 
 = \frac{|V_{ud}^2| + |V_{us}|^2}{f^2(\eps^{\mu e L}_{e \mu},\sum_{\alpha\beta P} |\eps^{\mu e P}_{\alpha\beta}|^2)} 
 = \frac{1}{f^2(\eps^{\mu e L}_{e \mu},\sum_{\alpha\beta P} |\eps^{\mu e P}_{\alpha\beta}|^2)} = 0.9999 \pm 0.0010
\end{equation}
at 1$\sigma$, where the CKM unitarity is inserted in the second
step. Again, this translates into a bound for $f(\eps^{\mu e L}_{e
  \mu},\sum_{\alpha\beta P} |\eps^{\mu e P}_{\alpha\beta}|^2)$, but
making the assumption of having only one non-zero $\eps$ at a time we
obtain:
\begin{eqnarray}
\label{eq:epsmueemuCKM}
|\real(\eps^{\mu eL}_{e\mu})| &<& 4.0\cdot 10^{-4}, \\
|\eps^{\mu eP}_{\alpha\beta}| &<& 0.030,
\end{eqnarray}
at the 90~\% confidence level. Notice that the bound of
\eq~(\ref{eq:epsmueemuCKM}) is slightly stronger than the one obtained
from the kinematic determination of the Fermi constant, but it relies
on one extra assumption, \ie, the unitarity of the CKM matrix.

On the other hand, if we assume that the NSI with quarks are
dominating, then the insertion of $G_\mu$ in place of $G_F$ is not
leading to any ambiguities. However, NSI of the form $\eps^{ud}$ will
contribute to the beta-decay rate, through which $V_{ud}$ is
extracted. Experimentally, only superallowed $0^+ \to 0^+$ decays are
considered, which means that the nuclear matrix element will have a
vector structure and, therefore, only the vector NSI combination will
contribute in the following way:
\begin{equation}
 \Gamma_\beta \propto G_F^2 |V_{ud}|^2 
 f(\eps^{ud V}_{e e},\sum_{\alpha} |\eps^{ud V}_{e \alpha}|^2).
\end{equation}
Since the Kaon decays are not affected by $\eps^{ud}$, these can be
used to extract $V_{ud}$ indirectly from the assumption of CKM
unitarity (\ie, $|V_{ud}|^2 = 1 - |V_{us}|^2$). The result of this
operation is $|V_{ud}|^2 = 0.94915 \pm 0.00086$~\cite{Amsler:2008zz},
which should be compared to the value of $|\tilde V_{ud}|^2=|V_{ud}|^2
f(\eps^{ud V}_{e e},\sum_{\alpha} |\eps^{ud V}_{e \alpha}|^2)$ derived
from beta decays $|\tilde V_{ud}|^2 = 0.94903 \pm
0.00055$~\cite{Amsler:2008zz}. Once again a truly model-independent
bound can only be extracted for the combination $f(\eps^{ud V}_{e
  e},\sum_{\alpha} |\eps^{ud V}_{e \alpha}|^2)$, but making the
assumption of taking one $\eps$ at a time we obtain:
\begin{eqnarray}
|\real(\eps^{ud V}_{ee})| &<& 0.00086,\\
|\eps^{ud V}_{e\alpha}| &<& 0.041.
\end{eqnarray}

Notice that, unlike the determination through the kinematic $G_F$, the
determination of the non-standard parameters $\eps^{\mu
  e}_{\alpha\beta}$ through CKM unitarity relies on the assumption
that the quark interactions are not affected, making the resulting
bounds slightly more model-dependent. On the other hand, if a given
model predicts both lepton ($\eps^{\mu e}_{\alpha\beta}$) and quark
($\eps^{u d}_{\alpha\beta}$) NSI simultaneously, the bounds on
$\eps^{\mu e}_{\alpha\beta}$ from the kinematic $G_F$ compared to muon
decay would still apply, while somewhat weaker bounds on $\eps^{u
  d}_{\alpha\beta}$ could still be derived after propagating the
errors derived on the former through the CKM unitarity relation.

\subsection{Bounds from pion processes}

For the quark charged-current NSI involving charged leptons other than
electrons, the universality tests stemming from the relative decay
rates of charged pions as well as that of taus into pions can be used
to set bounds. The squared and summed matrix element involving a
charged pion, a charged lepton and a neutrino is modified according to
\begin{equation}
 \sum_\beta |\mathcal M(\pi,\ell_\alpha,\nu_\beta)|^2 = 
 |\mathcal M(\pi,\ell_\alpha,\nu_\alpha)|^2 
f(\eps^{ud A}_{\alpha \alpha},\sum_{\beta} |\eps^{ud A}_{\alpha \beta}|^2).
\end{equation}
This modification is equivalent to violations of weak interaction
flavor universality identifying $g^2 f(\eps^{ud A}_{\alpha
  \alpha},\sum_{\beta} |\eps^{ud A}_{\alpha \beta}|^2) = g_\alpha^2$,
where $g_\alpha$ is the $W$ coupling to the lepton flavour
$\alpha$. Comparing the rates of $\pi \to e\nu$, $\pi \to \mu\nu$ and
$\tau \to \pi \nu$, bounds can be set on the ratios
$g_\alpha/g_\beta$. From \Ref~\cite{Loinaz:2004qc} we have
\begin{equation}
 \frac{g_\mu}{g_e} = 1.0021 \pm 0.0016 
  \quad {\rm and} \quad 
 \frac{g_\tau}{g_\mu} = 1.0030 \pm 0.0034\,
\end{equation}
at $1 \sigma$. Thus, if only one $\eps$ is considered at a time, we
obtain the following bounds at the $90~\%$ confidence level:
\begin{eqnarray}
\real(\eps^{ud A}_{\mu\mu}) &=& (2.1\pm 2.6)\cdot 10^{-3},\\
|\eps^{ud A}_{\mu \alpha}| &<& 0.078,\\
% from the first
\real(\eps^{ud A}_{\tau\tau}) &=& (3.0\pm5.5)\cdot 10^{-3},\\
|\eps^{ud A}_{\tau \alpha}| &<& 0.13,\\
% from the second. 
\real(\eps^{ud A}_{ee}) &=& (-2.1\pm 2.6)\cdot 10^{-3},\\
|\eps^{ud A}_{e \alpha}| &<& 0.045.
\end{eqnarray}
Notice that the bounds on $|\eps^{udA}_{e \alpha}|$ are more stringent
than the bounds on $|\eps^{udA}_{\mu\alpha}|$ because the offset of
the best-fit from the Standard Model expectation goes in the opposite
direction with respect to the effect of $|\eps^{udA}_{e \alpha}|$.

It is important to note that a model that predicts equal
$f(\eps^{udA}_{\alpha\alpha},\sum_{\beta}|\eps^{udA}_{\alpha\beta}|^2)$
for $\alpha = e, \mu, \tau$ cannot be bounded using this type of
argument, since it affects all of these decays in the same way and
universality is not violated.  However, if we only consider $\eps$ of
one chirality at a time, then this would imply that $\eps^{udP}_{\mu
  \alpha}$ and $\eps^{udP}_{\tau \alpha}$ share the stronger bounds
derived for $\eps^{udP}_{e \alpha}$ from the CKM unitarity.

In a similar fashion, we can use the universality test between the
$\mu \to e\nu\bar\nu$ and $\tau \to \mu\nu\bar\nu$ decays to constrain
the non-standard couplings $\eps^{\mu e}_{\alpha\beta}$. This
constraint is related to the lepton universality ratio $g_\tau/g_e =
1.0004\pm 0.0022$. Therefore, the inverse of this number is a
measurement of $\sqrt{f(\eps^{\mu e L}_{e \mu},\sum_{\alpha \beta P}
  |\eps^{\mu e P}_{\alpha \beta}|^2) }$, where we disregard possible
modifications of the tau decay which are not important for neutrino
oscillation experiments. The resulting bounds are:
\begin{eqnarray}
 \real(\eps^{\mu e}_{e\mu}) &=& (-0.4\pm3.5)\cdot10^{-3},\\ 
|\eps^{\mu e}_{\alpha\beta}| &<& 0.080.
\end{eqnarray}

\subsection{Bounds from oscillation experiments}

Production and detection NSI imply that a neutrino produced or
detected in association with a charged lepton will not necessarily
share its flavour. This means that flavour conversion is present
already at the interaction level and ``oscillations'' can occur at
zero distance. Indeed, in the presence of NSI,
\begin{equation}
P_{\alpha \beta}(L=0) 
\simeq |\eps^{u d A}_{\alpha \beta}|^2
\end{equation}
if the neutrino is produced through pion decays and
\begin{equation}
P_{e \alpha}(L=0) \simeq \sum_{\beta P} |\eps^{\mu e P}_{\alpha \beta}|^2
\quad \mathrm{as\ well\ as} \quad
P_{\mu \beta}(L=0) \simeq \sum_{\alpha P} |\eps^{\mu e P}_{\alpha \beta}|^2
\end{equation}
for neutrinos produced through muon decays. For the detection through
inverse beta decays the situation is a bit more involved since the
relative contributions of the different chiralities vary depending on
the energy regime due to the nuclear matrix elements.  Here we will
discuss the cases of very low ($E<1$ GeV) and very high ($E>10$ GeV)
energies.  In the first case the neutrino-nucleon cross section is
proportional to $(g_V^2 + 3 g_A^2)$, where $g_V=1$ and
$g_A=1.23$. This means that the vector and axial combinations of the
NSI that can mediate the processes will contribute incoherently with
those relative strengths to give:
\begin{equation}
P_{\alpha \beta}(L=0) 
\simeq \frac{1}{1+3g_A^2} (|\eps^{u d V}_{\beta \alpha}|^2 
+ 3g_A^2|\eps^{u d A}_{\beta \alpha}|^2).
\end{equation}
Notice that, if only one non-zero $\eps$ with definite chirality is present,
then
\begin{equation}
P_{\alpha \beta}(L=0) 
\simeq  |\eps^{u d P}_{\beta \alpha}|^2.
\end{equation}
We will make this assumption when we will summarise the bounds in the
last Section.  On the other hand, at very high energies, in the deep
inelastic scattering regime, the left-handed NSI contribute to the
neutrino cross-sections with a strength about twice that of the
right-handed, the actual factor being given by the ratio of the
neutrino and antineutrino cross sections at high energies for
an isoscalar target $r = \sigma_\nu/\sigma_{\bar\nu} \simeq 6.7/3.4 =
1.97$. We then obtain:
\begin{equation}
P_{\alpha \beta}(L=0) 
\simeq |\eps^{u d L}_{\beta\alpha}|^2 + \frac{1}{r}|\eps^{u d R}_{\beta \alpha}|^2.
\end{equation}
We can therefore use the very precise constraints on flavour
oscillations from experiments such as KARMEN~\cite{Eitel:2000by} and
NOMAD \cite{Astier:2001yj,Astier:2003gs}. Motivated by the large mass
hierarchies and small mixing angles observed in the quark sector,
these experiments explored neutrino oscillations at very short
baselines with high precision and no evidence of flavour change was
found. Both KARMEN and NOMAD produced neutrino beams from $\pi^+$
decays as well as the subsequent $\mu^+$ decays and detected them
through inverse beta decay. In the case of KARMEN the neutrinos were
produced via $\mu$ decays at rest, so that the neutrino energy was
always below $50$~MeV. On the other hand, NOMAD aimed at the
detection of $\nu_\tau$, so higher energies $\sim 20$ GeV were
exploited. Table \ref{tab-osc} contains a summary of the different
oscillation channels they explored and the bounds they imply for the
NSI parameters.
\begin{table}[t]
 \begin{center}
  \begin{tabular}{|c|c|c|c|}
   \hline
   {\bf Experiment} & {\bf Channel} &  {\bf Bounds} \\
   \hline
   \hline
   KARMEN   & $\bar\nu_\mu \to \bar\nu_e$ &
$\left|\eps^{\mu e P}_{\alpha e}\right|< 0.025$, 
$\left|\eps^{u d A}_{e \mu}\right| < 0.028$, 
$\left|\eps^{u d V}_{e \mu}\right| < 0.059$ \\ \hline
   NOMAD   & $\nu_\mu \to \nu_\tau$ & 
$\left|\eps^{u d A}_{\mu \tau}\right|< 0.013$, 
$\left|\eps^{u d L}_{\tau \mu}\right| < 0.013$, 
$\left|\eps^{u d R}_{\tau \mu}\right| < 0.018$  \\ \hline
   NOMAD   & $\nu_e \to \nu_\tau$ & 
$\left|\eps^{\mu e P}_{\alpha \tau}\right|< 0.087$, 
$\left|\eps^{u d L}_{\tau e}\right|<0.087$, 
$\left|\eps^{u d R}_{\tau e}\right|<0.12$ \\ \hline
   NOMAD   & $\nu_\mu \to \nu_e$ & 
$\left|\eps^{u d A}_{\mu e}\right|< 0.026$, 
$\left|\eps^{u d L}_{e \mu}\right|<0.026$, 
$\left|\eps^{u d R}_{e \mu}\right|<0.037$  \\ \hline
  \end{tabular}
  \caption{Bounds (90~\%~CL) from oscillations at zero distance. In
    each line, the first bound refers to production NSI, while the
    other two are for detection NSI.}
\label{tab-osc}
 \end{center}
\end{table}

\section{Loop Bounds}
\label{sec:loops}

\begin{figure}
  \begin{center}
    %% Non-mixing
     \begin{picture}(100,100)(0,0)
       \Text(50,100)[b]{(a)}
       \ArrowLine(10,10)(50,40)
       \Text(9,9)[tr]{$f$}
       \ArrowLine(50,40)(90,10)
       \Text(91,9)[tl]{$f$}
       \ArrowLine(10,90)(50,60)
       \Text(9,91)[br]{$\ell_\delta$}
       \Text(30,70)[tr]{$\nu_\delta$}
       \ArrowLine(50,60)(90,90)
       \Text(91,91)[bl]{$\ell_\gamma$}
       \Text(70,70)[tl]{$\nu_\gamma$}
       \GCirc(50,50){10}{.5}
       \Text(50,50)[c]{$\eps_{\gamma\delta}^{ff}$}
       \Photon(20,82.5)(80,82.5){2}{5.5}
       \Text(50,86)[b]{$W$}
     \end{picture}
     \hspace{0.1\textwidth}
    %% Mixing
     \begin{picture}(100,100)(0,0)
       \Text(50,100)[b]{(b)}
       \ArrowLine(10,10)(50,40)
       \Text(9,9)[tr]{$u$}
       \ArrowLine(50,40)(90,10)
       \Text(91,9)[tl]{$u$}
       \Text(65,28)[tr]{$d$}
       \ArrowLine(10,90)(50,60)
       \Text(9,91)[br]{$\mu$}
       \ArrowLine(50,60)(90,90)
       \Text(91,91)[bl]{$e$}
       \Text(70,77)[br]{$\nu_e$}
       \GCirc(50,50){10}{.5}
       \Text(50,50)[c]{$\eps_{\mu e}^{ud*}$}
       \Photon(80,17.5)(80,82.5){-2}{5.5}
       \Text(85,50)[l]{$W$}
     \end{picture}
    \caption{(a) The vanishing one-loop contribution to the mixing in
      the running between the matter NSI and the four-charged-fermion
      operator via $W$ exchange. (b) The non-vanishing one-loop
      contribution to the mixing in the running between the
      charged-current-like NSI and the operator inducing $\mu \to e$
      conversion in nuclei.}\label{fig:4fermion}
  \end{center}
\end{figure}
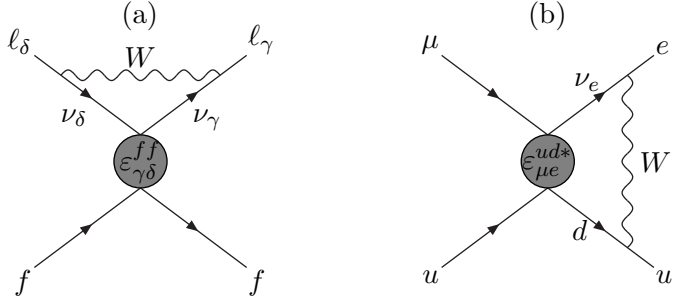

The tree level effects of neutrino NSI are difficult to constrain
since neutrino detection and flavour tagging is 
challenging. However, NSI may mix with four-charged-fermion operators
at the loop level inducing flavour-changing charged-lepton
interactions, for which strong bounds exist. In
\Ref~\cite{Biggio:2009kv} it was shown that, for a certain class of
diagrams (see \fig~\ref{fig:4fermion}a),
the logarithmic divergences that would indicate the mixing in the
running between NSI and four-charged-fermion operators
canceled. Therefore, only model-dependent finite contributions remain
and no model-independent bound can be derived through one-loop
considerations. We have checked that this is also the case for most
neutrino NSI at production and detection. There is, however, an
exception: the NSI parameter $\eps^{u d L}_{\mu e}$ mixes with the
operator that induces muon to electron conversion in nuclei through
the digram of \fig~\ref{fig:4fermion}b. The computation of this
diagram yields a logarithmic divergence:
\begin{equation}
\frac{3 \sqrt{2} G_F \alpha \eps^{ud L}_{\mu e}}{2 \pi s_w^2}\log\left(\frac{\Lambda}{M_W}\right) 
 [\bar u \gamma_\beta P_L u] [\bar \mu \gamma^\beta P_L e].
\label{loop}
\end{equation} 
Since the coefficient of this divergence can be interpreted as the
coefficient of the logarithmic running of this operator, we can
estimate the bound by assuming that $\log (\Lambda/M_W ) \simeq
1$. This gives a contribution to $\mu \to e$ conversion in nuclei of
the form (see, \eg, \Ref~\cite{Kitano:2002mt}):
\begin{equation}
{\rm R}(\mu^- \rightarrow e^-)
=\frac{m_\mu^5 (2V^{(p)} + V^{(n)})^2 |C|^2}
{\Gamma(\mu~ {\rm capture})},
\end{equation}
where $C$ is the coefficient of the operator in
\eq~(\ref{loop}). Using ${\rm R}(\mu^- \rightarrow e^-) < 7.0 \cdot
10^{-13}$ for conversion in Au~\cite{Amsler:2008zz} as well as
$V^{(p)}_{\rm Au} = 0.0974$ and $V^{(n)}_{\rm Au} =
0.146$~\cite{Kitano:2002mt}, a very strong bound on the NSI is
derived:
\begin{equation} 
|\eps^{ud L}_{\mu e}| < 1.8 \cdot 10^{-6}.
\end{equation}
We would like to remark that, also in this case, a quadratic
divergence is present. In principle, this contribution could dominate
over the logarithmic one, but its value is model-dependent and
reliable bounds cannot be derived from it. The contribution from the
logarithmic running could also be canceled, but only at a given
scale, which makes the resulting constraint more reliable.

\section{Neutral-current-like non-standard interactions}
\label{sec:nc}

For completeness, we will now also review the current status of the
bounds on NSI matter effects, or neutral-current-like NSI defined in
\eq~(\ref{matterNSI}). This type of NSI is the most extensively
studied in the literature, since it has been generally assumed that
the constraints on the charged-current-like NSI are much stronger. We
would like to stress that, in specific models, charged-current-like
and neutral-current-like processes are expected with similar
strengths~\cite{Antusch:2008tz}.

In most phenomenological studies the NSI parameters are reduced to the
effective parameters
\begin{equation}
 \eps_{\alpha\beta} = \sum_{f,P} \eps^{fP}_{\alpha\beta} \frac{n_f}{n_e},
\end{equation}
where $n_f$ is the number density of the fermion $f$. This is the
natural parameter in neutrino oscillation analyses since it
corresponds to the replacement
\begin{equation}
 H_{\rm matter} = V\mtrx{ccc}{1 & 0 & 0 \\ 0 & 0 & 0 \\ 0 & 0 & 0} \longrightarrow 
 V\left[\mtrx{ccc}{1 & 0 & 0 \\ 0 & 0 & 0 \\ 0 & 0 & 0} + 
   \mtrx{ccc}{\eps_{ee} & \eps_{e\mu} & \eps_{e\tau} \\ \eps_{e\mu}^* & \eps_{\mu\mu} & \eps_{\mu\tau} \\ \eps_{e\tau}^* & \eps_{\mu\tau}^* & \eps_{\tau\tau}}\right]
\end{equation}
in the matter interaction part of the neutrino flavour
evolution. Thus, assuming uncorrelated errors, the bounds on
$\eps_{\alpha\beta}$ could be approximated by
\begin{equation}
 \eps^{\oplus}_{\alpha\beta} \lesssim \left\{\sum_{P} 
      \left[\left(\eps_{\alpha\beta}^{eP}\right)^2 +
      \left(3\eps_{\alpha\beta}^{uP}\right)^2 +
      \left(3\eps_{\alpha\beta}^{dP}\right)^2\right]
    \right\}^{1/2}
\end{equation}
for neutral Earth-like matter with an equal number of neutrons and
protons and by
\begin{equation}
 \eps^{\odot}_{\alpha\beta} \lesssim \left\{\sum_{P} 
      \left[\left(\eps_{\alpha\beta}^{eP}\right)^2 +
      \left(2\eps_{\alpha\beta}^{uP}\right)^2 +
      \left(\eps_{\alpha\beta}^{dP}\right)^2\right]
    \right\}^{1/2}
\end{equation}
for neutral solar-like matter, consisting mostly of electrons and
protons. Using the bounds from
\Refs~\cite{Davidson:2003ha,Barranco:2005ps,Barranco:2007ej,Bolanos:2008km}, but
discarding the loop constraints on $\eps^{fP}_{e\mu}$
\cite{Biggio:2009kv}, the resulting bounds on the effective NSI
parameters would be
\begin{equation}
\label{eq:earth-sun}
 |\eps^\oplus_{\alpha\beta}| <
 \mtrx{ccc}{
  4.2 & 0.33 & 3.0 \\
  0.33 & 0.068 & 0.33 \\
  3.0 & 0.33 & 21
 }
 \quad {\rm and} \quad
 |\eps^\odot_{\alpha\beta}| <
 \mtrx{ccc}{
  2.5 & 0.21 & 1.7 \\
  0.21 & 0.046 & 0.21 \\
  1.7 & 0.21 & 9.0
 },
\end{equation}
respectively. Notice that atmospheric neutrino oscillations also
constrain the values of matter NSI through the relation
$\eps^\oplus_{\tau\tau} \simeq [|\eps^\oplus_{e\tau}|^2 \pm \mathcal
  O(0.1)]/(1+\eps^\oplus_{ee})$
\cite{Friedland:2004ah,Friedland:2005vy}. As long as
$1+\eps^\oplus_{ee}$ is not significantly smaller than one, this would
set a stronger bound $\eps^\oplus_{\tau\tau} \lesssim \mathcal O(10)$.

We want to stress the fact that the constraints on $\eps^e$, $\eps^u$
and $\eps^d$ have been derived under the assumption of taking one
non-zero $\eps$ at a time. Thus, the approach of combining them
together as in \eq~(\ref{eq:earth-sun}) is not fully consistent. For
this reason, in the compilation of all the results in the following
Section, the bounds will be quoted separately.

\section{Summary of results and conclusions}
\label{sec:results}

In order to easily overview our results, we here present the
constraints from the previous sections in tabularized format. In
\tab~\ref{tab:me} we present the different bounds available for
$\eps^{\mu e}_{\alpha\beta}$ while the bounds for
$\eps^{ud}_{\alpha\beta}$ are presented in \tab~\ref{tab:ud}. Taken
all together, the most stringent bounds available for both
charged-current-like and neutral-current-like NSI relevant for
terrestrial experiments are given by:
\begin{eqnarray}
 |\eps^{\mu e}_{\alpha\beta}| 
 &<&
 \mtrx{ccc}{0.025 & 0.030 & 0.030 \\ 0.025 & 0.030 & 0.030 \\ 0.025 & 0.030 & 0.030}, \\
\label{matrixud}
 |\eps^{ud}_{\alpha\beta}|
 &<&
 \mtrx{ccc}{0.041 & 0.025 & 0.041 \\ 
\begin{array}c
1.8 \cdot 10^{-6} \\[-.35cm]
0.026
\end{array}
 & 0.078 & 0.013 \\ 
\begin{array}c
0.087 \\[-.35cm]
0.12
\end{array} &
\begin{array}c
0.013 \\[-.35cm]
0.018
\end{array} &
0.13 },\\
 |\eps^e_{\alpha\beta}|
 &<&
  \mtrx{ccc}{
\begin{array}c
0.06 \\[-.35cm]
0.14
\end{array} &
0.10 & 
\begin{array}c
0.4 \\[-.35cm]
0.27
\end{array} \\
0.10 & 0.03 & 0.10 \\
\begin{array}c
0.4 \\[-.35cm]
0.27
\end{array} &
0.10 &
\begin{array}c
0.16 \\[-.35cm]
0.4
\end{array}},\\
 |\eps^u_{\alpha\beta}|
 &<&
  \mtrx{ccc}{
\begin{array}c
1.0 \\[-.35cm]
0.7
\end{array} &
0.05 & 0.5 \\
0.05 & 
\begin{array}c
0.003 \\[-.35cm]
0.008
\end{array} & 0.05 \\
0.5   & 0.05 &
\begin{array}c
1.4 \\[-.35cm]
3
\end{array}
 },\\
 |\eps^d_{\alpha\beta}|
 &<&
  \mtrx{ccc}{
 \begin{array}c
0.3 \\[-.35cm]
0.6
\end{array}  & 0.05 &  0.5 \\
 0.05  & 
\begin{array}c
0.003 \\[-.35cm]
0.015
\end{array} & 0.05 \\
 0.5  & 0.05 & 
\begin{array}c
1.1 \\[-.35cm]
6
\end{array}
 }.
\end{eqnarray}
Here, whenever two values are quoted, the upper value refers to
left-handed NSI and the lower to right-handed NSI. We would like to
stress that, before applying these constraints, the reader should
refer to the appropriate Sections in order to be aware of the
assumptions under which they were obtained.

\begin{table}
 \begin{center}
  \begin{tabular}{|c|c|c|c|c|}
   \hline
   {\bf $\bld{\eps^{\mu e}_{\alpha\beta}}$} 
 & {\bf Kin.~$\bld{G_F}$} $(L,R)$ 
 & {\bf CKM unit.} $(V)$ 
 & {\bf Lept.~univ.} $(A)$ 
 & {\bf Oscillation} $(L,R)$  \\
   \hline
   \hline
   $\eps^{\mu e}_{ee}$    & $< 0.030$ & $< 0.030$ & $< 0.080$ & $< 0.025$ \\
%   & & & & (KARMEN) \\
   \hline
   $\eps^{\mu e}_{e\mu}$  & $(-1.4\pm1.4)\cdot10^{-3}$($\mathbb R$,$L$) & $< 4\cdot 10^{-4}$($\mathbb R$) & $(-0.4\pm3.5)\cdot10^{-3}$($\mathbb R$) & - \\
                          & $< 0.030$ & $< 0.030$ & $< 0.080$ & \\
   \hline
   $\eps^{\mu e}_{e\tau}$ & $< 0.030$ & $< 0.030$ & $< 0.080$ & $< 0.087$ \\
%   & & & & (NOMAD) \\
   \hline
   $\eps^{\mu e}_{\mu e}$ & $< 0.030$ & $< 0.030$ & $< 0.080$ & $< 0.025$ \\
%   & & & & (KARMEN) \\
   \hline
   $\eps^{\mu e}_{\mu \mu}$ & $< 0.030$ & $< 0.030$ & $< 0.080$ & - \\
   \hline
   $\eps^{\mu e}_{\mu\tau}$ & $< 0.030$ & $< 0.030$ & $< 0.080$ & $< 0.087$ \\
%   & & & & (NOMAD) \\
   \hline
   $\eps^{\mu e}_{\tau e}$ & $< 0.030$ & $< 0.030$ & $< 0.080$ & $< 0.025$ \\
%   & & & & (KARMEN) \\
   \hline
   $\eps^{\mu e}_{\tau \mu}$ & $< 0.030$ & $< 0.030$ & $< 0.080$ & - \\
   \hline
   $\eps^{\mu e}_{\tau \tau}$ & $< 0.030$ & $< 0.030$ & $< 0.080$ & $< 0.087$ \\
%   & & & & (NOMAD) \\
   \hline
  \end{tabular}
  \caption{Bounds (90~\%~CL) on the purely leptonic
    charged-current-like NSI $\eps^{\mu e}_{\alpha\beta}$, relevant to
    the neutrino production through muon decay, \eg, at a Neutrino
    Factory. The letters $L, R, V, A$ refer to the chirality of the
    $\eps$ which is actually bounded, while $\mathbb R$ stands for the
    real part of the element only. See the text for details.}
  \label{tab:me}
 \end{center}
\end{table}

\begin{table}
 \begin{center}
  \begin{tabular}{|c|c|c|c|c|}
   \hline
   $\bld{\eps^{ud}_{\alpha\beta}}$ 
 & {\bf CKM unit.} $(V)$ 
 & {\bf Lept.~univ.} $(A)$
 & {\bf Oscillation}  
 & {\bf Loop} $(L)$\\
   \hline
   \hline
   $\eps^{ud}_{ee}$    & $< 8.6\cdot 10^{-4}$($\mathbb R$) & $(-2.1\pm 2.6)\cdot 10^{-3}$($\mathbb R$) & - & - \\
                       & $< 0.041$ & $< 0.045$ & & \\
   \hline
   $\eps^{ud}_{e\mu}$  & $< 0.041$ & $< 0.045$ & $< 0.028 (A)$ & - \\
%   & & & (KARMEN) & \\
    &&&$< 0.059 (V)$ &  \\
    &&&$< 0.026 (L)$ &  \\
    &&&$< 0.037 (R)$ &  \\
   \hline
   $\eps^{ud}_{e\tau}$ & $< 0.041$ & $< 0.045$ & - & - \\
   \hline
   $\eps^{ud}_{\mu e}$ & - & $< 0.078$ & $< 0.026 (A)$ & $< 1.8\cdot 10^{-6}$\\
%   & & & (NOMAD) & \\
   \hline
   $\eps^{ud}_{\mu \mu}$ & - & $(2.1\pm 2.6)\cdot 10^{-3}$($\mathbb R$) & - & - \\
   & & $< 0.078$ & & \\
   \hline
   $\eps^{ud}_{\mu\tau}$  & - & $< 0.078$ & $< 0.013 (A)$ & - \\
%   & & & (NOMAD) & \\
   \hline
   $\eps^{ud}_{\tau e}$ & - & $< 0.13$ & $< 0.087 (L)$ & - \\
   & & & $< 0.12 (R)$ & - \\
%   & & & (NOMAD) & \\
   \hline
   $\eps^{ud}_{\tau \mu}$ & - & $< 0.13$ & $< 0.013 (L)$ & - \\
   & & & $< 0.018 (R)$ & - \\
%   & & & (NOMAD) & \\
   \hline
   $\eps^{ud}_{\tau \tau}$ & - & $(3.0\pm5.5)\cdot 10^{-3}$($\mathbb R$) & - & - \\
   & & $< 0.13$ & & \\
   \hline
  \end{tabular}
  \caption{Bounds (90~\%~CL) on the quark charged-current-like NSI
    $\eps^{ud}_{\alpha\beta}$, relevant to the neutrino production
    through hadron decays as well as detection processes. The letters
    $L, R, V, A$ refer to the chirality of the $\eps$ which is
    actually bounded, while $\mathbb R$ stands for the real part of
    the element only. See the text for details.}
  \label{tab:ud}
 \end{center}
\end{table}

To summarise, we have presented the model-independent bounds that can
be derived for various types of NSI. Since the neutral-current-like
NSI have been studied extensively in the literature and the bounds on
these are fairly well known, we have just summarised these results and
concentrated on the charged-current-like NSI, which usually are simply
considered to be very strongly bounded, although no model-independent
analysis has been readily available. The result of our analysis is
that the charged-current-like NSI, which are of interest mostly for
their impact on neutrino production and detection, are generally
bounded by numbers of $\mathcal O(10^{-2})$--$\mathcal O(10^{-1})$,
except for the very strong loop bound on $\eps^{udL}_{\mu e}$ due to
the operator mixing inducing $\mu \to e$ conversion in nuclei. We find
that these bounds are about one order of magnitude stronger than the
bounds on the neutral-current-like NSI. We therefore argue that production
and detection NSI should not be neglected with respect to matter NSI,
especially taking into account that, in most realisations, both kinds
of NSI are induced with similar strengths.  Moreover, NSI saturating
the bounds derived here will be within the sensitivity reach of
planned neutrino oscillation experiments. However, as discussed in the
introduction, most models leading to NSI generally affect other
processes and therefore stronger bounds than the ones derived here
apply.

\begin{acknowledgments}
The authors are grateful to Hisakazu Minakata for encouragement and to
Davide Meloni and Michele Papucci for useful discussions.  This work
was supported by the Swedish Research Council (Vetenskapsr{\aa}det),
contract no.~623-2007-8066 [M.B.]. M.B.\ and E.F.M.\ would also like
to thank NORDITA for warm hospitality during the programme
``Astroparticle Physics - A Pathfinder to New Physics'' during which
parts of this work was performed.  The authors also acknowledge
support by the DFG cluster of excellence ``Origin and Structure of the
Universe''. E.F.M.\ also acknowledges support from the European
Community under the European Commission Framework Programme 7 Design
Study: EUROnu, Project Number 212372.
\end{acknowledgments}

\end{document}